\documentclass{article}
\usepackage{graphicx}
\usepackage{listings}
\usepackage{wrapfig}
\usepackage{subfigure}
\usepackage{listings}
\usepackage{enumitem}
\usepackage{times}
\usepackage{todonotes}
\usepackage{soul}
\usepackage{url}
\usepackage{amsmath}
\usepackage{multirow}
\usepackage{url}
\usepackage{float}
\usepackage{caption}
\usepackage{setspace}
\usepackage{floatflt}
\usepackage{times}
\usepackage{graphicx}
\usepackage{listings}
\usepackage{wrapfig}
\usepackage{subfigure}
\usepackage{listings}
\usepackage{enumitem}
\usepackage{times}
\usepackage{todonotes}
\usepackage{soul}
\usepackage{url}
\usepackage{amsmath}
\usepackage{multirow}
\usepackage{adjustbox,lipsum}
\makeatletter
\renewcommand\part{%
  \if@openright
    \cleardoublepage
  \else
    \clearpage
  \fi
  \thispagestyle{empty}%
  \if@twocolumn
    \onecolumn
    \@tempswatrue
  \else
    \@tempswafalse
  \fi
  \null\vfil
  \secdef\@part\@spart}
\makeatother
\usepackage[numbers,sort]{natbib}
\usepackage{authblk}
\providecommand{\keywords}[1]{\textbf{\textit{Index terms---}} #1}
\usepackage{url}
\usepackage{float}
\usepackage{caption}
\usepackage{setspace}
\usepackage{floatflt}
\usepackage{times}
\usepackage{mwe,tikz}
\usepackage[percent]{overpic}
\usepackage{tikz}
\usepackage{tkz-tab}
\usepackage{setspace}
\usepackage{floatflt}

\begin{document}

\title{Characterization of behavioral patterns exploiting description of geographical areas }

%
%

\author[1]{Zolzaya Dashdorj}
\author[2]{Stanislav Sobolevsky}

\affil[1]{University of Trento (Via Sommarive, 9 Povo, TN, Italy), and SKIL LAB - Telecom Italia, and DKM - Fondazione Bruno Kessler, and}
\affil[ ]{SICT - Mongolian University of Science and Technology (Bayanzurkh district 22th khoroo, UB, Mongolia) }
\affil[ ]{\url{dashdorj@disi.unitn.it}}
\affil[2]{Massachusetts Institute of Technology (MIT 77 Massachusetts Avenue Cambridge, MA, USA) and}
\affil[ ]{New York University (1 MetroTech Center, Brooklyn, NY) }
\affil[ ]{\url{stanly@mit.edu}}

\date{}
\maketitle

\begin{abstract}

The enormous amount of recently available mobile phone data is providing unprecedented direct measurements of human behavior. Early recognition and prediction of behavioral patterns are of great importance in many societal applications like urban planning, transportation optimization, and health-care. Understanding the relationships between human behaviors and location's context is an emerging interest for understanding human-environmental dynamics. Growing availability of Web 2.0, i.e. the increasing amount of websites with mainly user created content and social platforms opens up an opportunity to study such location's contexts. This paper investigates relationships existing between human behavior and location context, by analyzing log mobile phone data records. First an advanced approach to categorize areas in a city based on the presence and distribution of categories of human activity (e.g., eating, working, and shopping) found across the areas, is proposed. The proposed classification is then evaluated through its comparison with the patterns of temporal variation of mobile phone activity and applying machine learning techniques to predict a timeline type of communication activity in a given location based on the knowledge of the obtained category vs. land-use type of the locations areas. The proposed classification turns out to be more consistent with the temporal variation of human communication activity, being a better predictor for those compared to the official land use classification.
\end{abstract}

%

\section{Introduction}\label{introduction}

Recent extensive penetration of digital technologies into everyday life have enabled creation and collection of vast amounts of data related to different types of human activity. When available for research purposes this creates an unprecedented opportunity for understanding human society directly from it's digital traces. There is an impressive amount of papers leveraging such data for studying human behavior, including mobile phone records \cite{ratti2006mlu, calabrese2006real, girardin2008digital, quercia2010rse, ratti2010redrawing}, vehicle GPS traces \cite{santi2013taxi, kang2013exploring},  social media posts \cite{java2007we, hawelka2014, paldino2015flickr} and bank card transactions \cite{sobolevsky2014mining, sobolevsky2014money}. With the growing mobile phone data records, environment modeling can be designed and simulated for understanding human dynamics and correlations between human behaviors and environments. Environment modeling is important for a number of applications such as navigation systems, emergency responses, and urban planning. Researchers noticed that type of the area defined through official land-use is strongly related with the timeline of human activity \cite{RePEc:pio:envirb:v:36:y:2009:i:5:p:824-836,DBLP:conf/socialcom/Frias-MartinezSHF12,Phithakkitnukoon:2010:AMI:1881331.1881336,
Wakamiya:2011:UAC:2008664.2008674,DBLP:conf/icwsm/NoulasSMP11a}. But those sources of literature do not provide extensive analyses on categorical profile of the geographical areas. This limits the understanding of the dependency of human behaviors from geographical areas. Our analysis confirms this relation, however we show that land-use by itself might be not enough, while categorical profile of the area defined based on OSM provides a better prediction for the activity timeline. For example, even within the same land-use category, timelines of activity still vary depending on the categorical profile. In this paper, different from these works, we start from clustering the entire city based on area profiles, that are a set of human activities associated with a geographical location, showing that those activities have different area types in terms of the timelines of mobile phone communication activity. Further we show that even the areas of the same land-use, which is formally defined by land-use management organizations, might have different clusters based on points of interest (POIs). But those clustered areas are still different in terms of the timelines. This will contribute to other works showing that not only the land-use matters for human activity. This paper uses mobile phone data records to determine the relationship between human behaviors and geographic area context \cite{Dashdorj:2015:ISC:2800835.2801625}. We present a series of experimental results by comparing the clustering algorithms aiming at answering the following questions: 1). To what extent can geographical types explain human behaviors in a city, 2). What is the relationship between human behaviors and geographical area profiles? We demonstrate our approach to predict area profiles based on the timelines of mobile phone communication activities or vice versa: to predict the timelines from area profiles. We validate our approach using a real dataset of mobile phone and geographic data of Milan, Italy. Our area clustering techniques improve the overall accuracy of the baseline to 64.89\%. Our result shows that land-uses in city planning are not necessarily well defined that an area type is better defined with one type of human activity. But growing and development of city structures enable various types of activities that are present in one geographical area. So this type of analysis and its application is important for determining robust land-uses for city planning. Also the hidden patterns and unknown correlations can be observed comparing the mobile phone timelines in relevant areas. The result of this work is potentially useful to improve the classifications of human behaviors for better understanding of human dynamics in real-life social phenomena and to provide a decision support for stakeholders in areas, such as urban city, transport planning, tourism and events analysis, emergency response, health improvement, community understanding, and economic indicators. The paper is structured as follows Section \ref{sec:datasource} introduces the data sources we use in this research and the data-processing performed. The methodology is described in Section \ref{sec:approach}. We present and discuss the experimental results in Section \ref{sec:experiment}. Finally, we summarize the discussions in Section \ref{sec:conclusion}.

\section{Related Works}

Human behavior is influenced by many contextual factors and their change, for instance, snow fall, hurricane, and festival concerts. There are number of research activities that shed new light on the influence of such contextual factors on social relationships and how mobile phone data can be used to investigate the influence of context factors on social dynamics. Researchers~\cite{DBLP:journals/corr/abs-1106-0560, Phithakkitnukoon:2010:AMI:1881331.1881336, Calabrese:2010:GTA:2166616.2166619,Furletti:2012:IUP:2346496.2346500} use an additional information about context factors like social events, geographical location, weather condition, etc in order to study the relationship between human behaviors and such context factors. This is always as successful as the quality of the context factors. The combination of some meteorological variables, such as air temperature, solar radiation, relative humidity,  can effect people's comfort conditions in outdoor urban spaces~\cite{Stathopoulos2004297}, poor or extreme weather conditions influence peoples physical activity~\cite{Tucker2007909}. Q. Wang et al.~\cite{10.1371/journal.pone.0112608} exhibited high resilience, human mobility data obtained in steady states can possibly predict the perturbation state. The results demonstrate that human movement trajectories experienced significant perturbations during hurricanes during/after the Hurricane Sandy in 2012. Sagl et al.~\cite{DBLP:conf/icsdm/SaglBRB11} introduced an approach to provide additional insights in some interactions between people and weather. Weather can be seen as a higher-level phenomenon, a conglomerate that comprises several meteorological variables including air temperature, rainfall, air pressure, relative humidity, solar radiation, wind direction and speed, etc. The approach has been significantly extended to a more advanced context-aware analysis in \cite{s120709800}. Phithakkitnukoon et al. \cite{Phithakkitnukoon:2010:AMI:1881331.1881336} used POIs to enrich geographical areas. The areas are connected to a main activity (one of the four types of activities investigated) considering the category of POIs located within it. To determine groups, that have similar activity patterns, each mobile user's trajectory is labeled with human activities using Bayes Theorem in each time-slot of a day for extracting daily activity patterns of the users. The study shows that daily activity patterns are strongly correlated to a certain type of geographic area that shares a common characteristic context. Similar to this research idea, social networks \cite{Wakamiya:2011:UAC:2008664.2008674} have been taken into account to discover activity patterns of individuals. Noulas et al. \cite{DBLP:conf/icwsm/NoulasSMP11a} proposed an approach for modelling and characterization of geographic areas based on a number of user check-ins and a set of eights type of general (human) activity categories in Foursquare. A Cosine similarity metric is used to measure the similarity of geographical areas. A Spectral Clustering algorithm together with K-Means clustering is applied to identify an area type. The area profiles enables us to understand  groups of individuals who have similar activity patterns. Soto and Frias-Martinez et al. \cite{Soto11robustland} studied mobile phone data records to characterize geographical areas with well defined human activities, by using the Fuzzy C-Means clustering algorithm. The result indicated that five different land-uses can be identified and their representation was validated with their geographical localization by the domain experts. Frias-Martinez et al. \cite{DBLP:conf/socialcom/Frias-MartinezSHF12} also studied geolocated tweets to characterize urban landscapes using a complimentary source of land-use and landmark information. The authors focused on determining the land-uses in a specific urban area based on tweeting patterns, and identification of POIs in areas with high tweeting activity. Differently, Yuang et al. \cite{conf/giscience/YuanR12} proposed to classify urban areas based on their mobility patterns by measuring the similarity between time-series using the Dynamic Time Warping (DTW) algoritm. Some areas focus on understanding urban dynamics including dense area detection and their evolution over time \cite{Vieira:2010:CDU:1906497.1907403,conf/icde/NiR07}. Moreover, \cite{clioandris:ratti2006mobile,Fujisaka:2010:EUC:1734583.1734588,soto11robust} analyzed mobile phone data to characterize urban systems. More spatial clustering approaches ( Han \& Kamber \cite{citeulike:2855241}) could group similar spatial objects into classes, such as k-means, k-medoids, and Self Organizing Map. They have been also used for performing effective and efficient clustering. In this research, we use spectral clustering with eigengap heuristic followed by k-means clustering. Calabrese et al. \cite{RePEc:pio:envirb:v:36:y:2009:i:5:p:824-836} and also \cite{journals/corr/PeiSRSZ13,DBLP:journals/corr/GrauwinSMGR14} used eigengap heuristic for clustering urban land-uses. In many works \cite{pei2014new,RePEc:eee:phsmap:v:390:y:2011:i:5:p:929-942,10.1371/journal.pone.0081707,tnhh:reades:census,clioandris:ratti2006mobile,Becker_atale} the authors analyzed mobile phone data activity timelines to interpret land-use type. T. Pei et al.~\cite{pei2014new} analyzed the correlation between urban land-use information and mobile phone data. The author constructed a vector of aggregated mobile phone data to characterize land-use types composed of two aspects: the normalized hourly call volume and the total call volume. A semi-supervised fuzzy c-means clustering approach is then applied to infer the land-use types. The method is validated using mobile phone data collected in Singapore. land-use is determined with a detection rate of 58.03\%. An analysis of the land-use classification results shows that the detection rate decreases as the heterogeneity of land-use increases, and increases as the density of cell phone towers increases. F. Girardin et al.~\cite{Girardin_quantifyingurban} analyzed aggregate mobile phone data records in New York City to explore the capacity to quantify the evolution of the attractiveness of urban space and the impact of a public event on the distribution of visitors and on the evolution of the attractiveness of the points of interest in proximity.

\section{Collecting and Pre-Processing the data}\label{sec:datasource}

We use two types of datasources for this experiment; 1) POIs from available geographical maps, Openstreetmap 2) Mobile phone network data (sms, internet, call, etc) generated by the largest operator company in Italy. The mobile phone traffic data is provided in a spatial grid, the rectangular grid of 100 x 100 square of dimension 235 $m$ x 235 $m$. We use the grid as our default configuration for collecting human activity distribution and mobile network traffic activity distribution. 


\subsection{Point of Interests extracted from Openstreetmap}

\begin{figure*}[htb!]
       \centering
       \subfigure[The location size (aggregation radius * 2) distribution]
       {\includegraphics[scale=0.3]{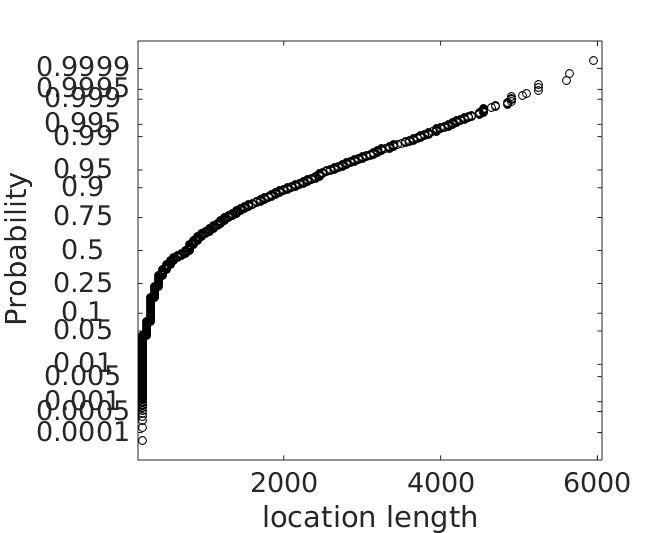}
         \label{gra:location size distribution}}
       \quad
        \subfigure[Human activity relevant POI distribution considering aggregation radius]
       {\includegraphics[scale=0.28]{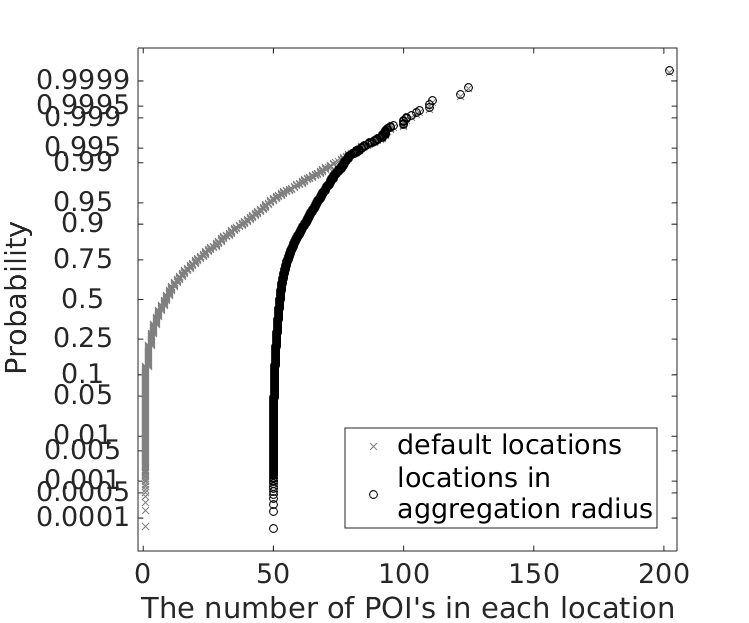}
       \label{gra:POI Human activity distribution nearby areas Milan}}
       \caption{The distributions of POIs and human activities across locations}
\end{figure*}

In \cite{DBLP:conf/mum/DashdorjSAL13,Dashdorj:2014:HAR:2675316.2675321,Zolzaya:DC:2013,Zolzaya:Poster:2013}, one of the key elements in the contextual description of geographical regions is the point of interest (POI) (e.g. restaurants, ATMs, and bus stops) that populates an area. A POI is a good proxy for predicting the content of human activities in each area that was well evaluated in \cite{Zolzaya2015}. Employing a model proposed in \cite{Zolzaya2015}, a set of human activities likely to be performed in a given geographical area, can be identified in terms of POI distribution. This allows us to create area profiles of geographical locations in order to provide semantic (high level) descriptions to mobile phone data records in Milan. For example, a person looking for food if the phone call is located close to a restaurant. We exploit the given spatial grid to enrich the locations with POIs from  open and free geographic information, Openstreetmap (OSM)\footnote{http:\///www.openstreetmap.org}. We collected in total 552,133 POIs that refined into 158,797 activity relevant POIs. To have a sufficient number and diversity of POIs in each location, we consider the nearby areas for estimating the likelihood of human activities. The nearby areas are the intersected locations within the aggregation radius of the centroid point at each location. The aggregation radius is configured differently in each location, which satisfies the need for the total number of POIs in such intersected locations to be above the threshold \textit{h}, see Figure \ref{gra:POI Human activity distribution nearby areas Milan} and \ref{gra:location size distribution} where each location at least $h$=50 POIs in the intersected locations. Across locations, the min, median, and max number of POIs are 50, 53, and 202. 

In order to build area profiles of each location, a $n$ x $m$ dimensional matrix $A_{n,m}$ is defined for each location $n \in \{1,..,10000\}$. Each element $A_{n,m}$ contains the weight of activity categories $m$ in location \textit{n} where the $m \in \{${eating, educational, entertainment, health, outdoor, residential, shopping, sporting, traveling, working}$\}$, with the total number of 10 measurements of human activities per each location. The weight of each category of activities are estimated by the HRBModel which allows us to generate a certain weight for human activities that is proportional to the weight of relevant POIs located in each location. The weight of POIs in a given location, is estimated by the following equation of $tf-idf(f, l) = \frac{N(f, l)}{\mathop{\rm argmax}\limits_w \{N(w, l) : w \in l\}} * \log \frac{|L| }{ |\{l \in L: f \in l\}|}$, where $f$ is a given POI; $f \in F$, $F$=\{building, hospital, supermarket,...\} and $l$ is a given location; $l \in L$, $L$=\{location1, location2, location3,...\}, $N(f, l)$ is the occurrence of POI \textit{f} and its appearance in location \textit{l} \ and $\mathop{\rm argmax}\limits_w \{N(w, l) : w \in l\}$ is the maximum occurrence of all the POIs in location \textit{l}, $|L|$ is the number of all locations, $|\{l \in L: f \in l\}|$ is the number of locations where POI \textit{f} appears.

\begin{figure}[htb!]
  \centering
  \includegraphics[scale=0.4]{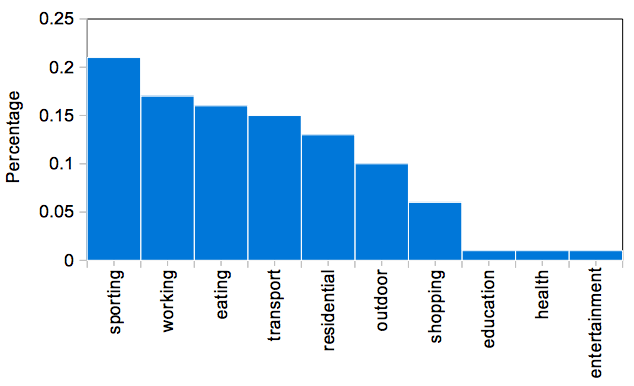}
  \caption{The activity distribution in Milan}
  \label{gra:activity distribution Milan}
\end{figure}

The activity distribution in Milan area is shown in Figure \ref{gra:activity distribution Milan}. The sporting, working, eating and transportation types of activities are mainly performed in the city.

\subsection{Mobile phone network traffic}

In this work, we used a dataset from ``BigDataChallenge''\footnote{http:\//\//www.telecomitalia.com\//tit\//it\//bigdatachallenge.html} organized by Telecom Italia. The dataset is the result of a computation over the Call Detail Records (CDRs) generated by the Telecom Italia cellular network within Milan. The dataset covers 1 month with 180 million mobile network events in November, 2014 as November is a normal month without any particular events organized in Milan. The CDRs log the user activity for billing purposes and network management. There are many types of CDRs, for the generation of this dataset we considered those related to the following activities: square id (the id of the square that is part of the Milan GRID which contains spatially aggregated urban areas), time interval (an aggregate time), received SMS (a CDR is generated each time a user receives an SMS), sent SMS (a CDR is generated each time a user sends an SMS), incoming Calls (a CDR is generated each time a user receives a call), outgoing Calls (a CDR is generated each time a user issues a call), internet (a CDR is generate each time, a user starts an internet connection, or a user ends an internet connection).


By aggregating the aforementioned records, this dataset was created that provides mobile phone communication activities across locations. The call, sms and internet connection activity logs are collected in each square of the spatial grid for Milan urban area. The activity measurements are obtained by temporally aggregating CDRs in time-slots of ten minutes. But the temporal variations make the comparison of human behaviors more difficult. The standard approach to account for temporal variations in human behavior is to divide time into coarse grained time-slots. In Farrahi and Gatica-Perez et al. \cite{farrahi:acmmm:2008}, the following eight coarse-grained time-slots are introduced: [00-7:00 am., 7:00-9:00 am., 9:00-11:00 am., 11:00 am.-2:00 pm., 2:00-5:00 pm., 5:00-7:00 pm., 7:00-9:00 pm., and 9:00 pm.-00 am.]. Here, we aggregate the mobile phone network data in such coarse-grained time-slots to extract the pattern of 1 month network traffic volume in each location. For each location, we then aggregated the total number of call (outgoing and incoming call without considering a country code), and sms (incoming and outgoing), internet  activity for each of those eight time-slots. Such time-slot based timelines can give us actual patterns of mobile network traffic activity.

\begin{figure*}[htb!]
\centering
  \includegraphics[scale=1,width=12cm]{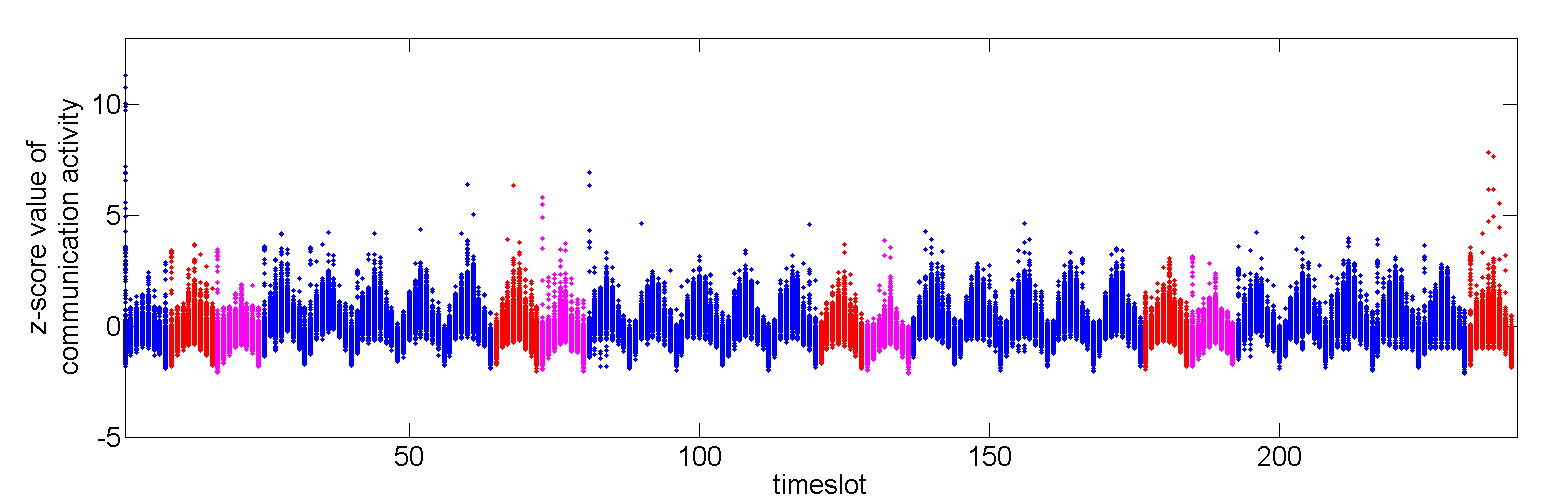}
   \caption{The timelines for each time-slot of day normalized by z-score in each location (weekday-blue, saturday-red, sunday-pink)}
   \label{gra:timelines2}
\end{figure*}

Then the dataset reduced to 2.4 million CDR each of which consists of the followings: square id, day of month, time-slot, and total number of mobile network traffic activity. We build a $n$ x $p$ x $d$ dimensional matrix $T_{n,p,d}$ to collect a mobile phone traffic activity timeline, where $n$ is the number of locations in [1,10000], $p$ is the time-slot divisions of the day [1,8] and $d$ is the day in [1,31]. To identify timeline patterns among those locations, we performed a normalization for the timelines based on z-score which transforms the timeline into the output vector with mean $\mu$=0 while standard deviation $\sigma$ is negative if it is below the mean or positive if it is above the mean. The normalized timelines by day are visualized in Figure \ref{gra:timelines2} which show a stable communication activity within the month. For this transformation, we used $T'_{i,j,k} = \dfrac{T_{i,j,k} - \mu_i }{\sigma_i}, i \in n, j \in p, k \in d$, where $\mu_i$ is the average value of the mobile phone activity traffic in location $i$, $\sigma_i$ is the standard deviation of the mobile phone activity traffic in location $i$.


\section{The Approach}\label{sec:approach}
We present our methodology for identifying the relation between geographical locations and human behaviors. Our methodology is divided into two phases: 1) clustering approaches for inferring categorical area types in terms of geographical area profiles 2) classification approaches for validating the observed area types by mobile phone data records. Clustering techniques are mostly unsupervised methods that can be used to organize data into groups based on similarities among the individual data items. We use the spectral clustering algorithm which makes use of the spectrum (eigenvalues) of the similarity matrix of the data to perform dimensionality reduction before clustering in fewer dimensions. The similarity matrix is provided as an input and consists of a quantitative assessment of the relative similarity of each pair of points in the dataset.

We define a vector space model that contains a set of vectors corresponding to areas. Relevance between areas is a similarity comparison of the deviation of angles between each area vector. The similarity between the areas is calculated by the cosine similarity metric by estimating the deviation of angles among area vectors. For example, the similarity between area $l_1$ and $l_2$ would be $\cos{\theta}_{l_1,l_2} = \frac{\mathbf{l_1} \cdot \mathbf{l_2}}{\left\| \mathbf{l_2} \right\| \left \| \mathbf{l_1} \right\|}$ where $l_i$ denotes the area or the features associated to the areas. We denote each area $l_i$ with a set of corresponding features associated with a weight measure $j$. Having the estimation of similarity between the areas, we can now create a similarity graph described as the weight matrix $W$ generated by the cosine similarity metrics and the diagonal degree matrix $D$ is utilized by the spectral clustering algorithm which is the one of the most popular modern clustering methods and performs better than traditional clustering algorithms. We create the adjacency matrix $A$ of the similarity graph and graph Laplacian $LA$,  $LA=D-A$ (given by normalized graph Laplacian $LA_n=D^{-1/2}LAD^{-1/2}$). Based on eigengap heuristic \cite{DBLP:journals/corr/abs-0711-0189}, we identify the number of clusters by k-nearest neighbor to observe in our dataset as $k = argmax_{i}(\lambda_{i+1}-\lambda_{i})$ where $\lambda_i \in \{l_1, l_2, l_3,.., l_n\}$ denotes the eigenvalues of $l_n$ in the ascending order. Finally, we easily detect the effective clusters (area profiles) $S_1, S_2, S_3, ..., S_k$ from the first $k$ eigenvectors identified by the k-means algorithms.

We investigate the relation between geographical locations and human behaviors based on categorical area types. To do that, we use supervised learning algorithms to predict area profile of a given area if we train a classification model with training data, which are the timelines labeled with area types. In supervised learning, each observation has a corresponding response or label. Classification models learn to predict a discrete class given new predictor data. We use several of classifiers for learning and prediction. We prepare a test set for testing classification models by k-fold cross validation method.  

%

\section{Experiments and Results}\label{sec:experiment}

In this section, we demonstrate the identification of the relationships between locations and human behaviors in terms of two types of features in each location: 1) location contexts: categories of human activity estimated through types of available POI 2) mobile communication activity timeline: mobile communication activity in time-series of coarse grained time-slots. In other words, the extent to which human behaviors depend on geographical area types. To identify and quantify these dependencies, we perform two types of validations: 1) observed area type we defined vs human behavior 2) land-use type defined formally vs human behavior by estimating the correlations and prediction algorithms. 

\begin{figure}[htb!]
       \centering
       \includegraphics[scale=0.3]{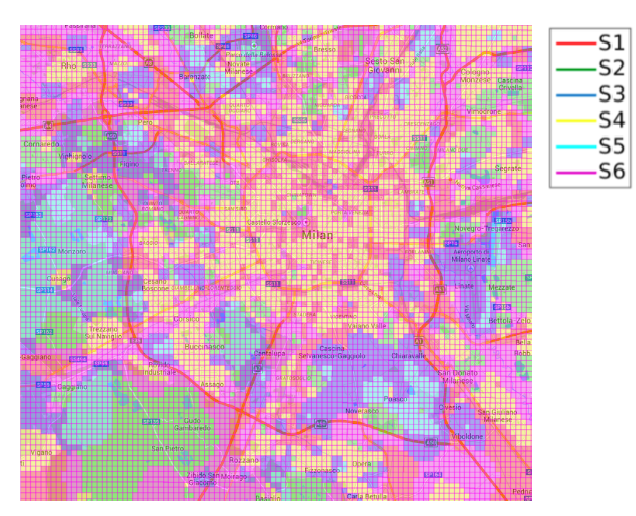}
       \caption{Observed area types of the geographical area of Milan based on the area profiles, $k$=6, where $S1$ is red, $S2$ is lime, $S3$ is blue, $S4$ is yellow, $S5$ is cyan/aqua, and $S6$ is magenta/fuchsia}
       \label{gra:clustering high level k =6}
\end{figure}

\subsection{Observed area type vs human behavior}

We first check the two datasets can be clustered or randomly distributed using Hopkins statistic, $H=\frac{\sum_{i=1}^{n} y_i}{\sum_{i=1}^{n} x_i+\sum_{i=1}^{n} y_i}$. The distance between element $p_i$ and its nearest neighbor in dataset $D$ is $x_{i}=\min_{v \in D}\left \{ dist(p_i,v) \right \}$ and the distance between element $q_i$ and its nearest neighbor in $D-{q_i}$ is $y_{i}=\min_{v \in D,v\neq q_i}\left \{ dist(q_i,v) \right \}$. The Hopkins statistic for the location context dataset is 0.02 and the mobile communication timeline is 0.04 that indicates that the datasets are highly clustered and regularly distributed. So we then analyze the correlations of location context and mobile phone communication timeline in order to understand if humans are attracted to location contexts through the area types (i.e., shopping, woking, and studying). To validate such relationship, we start with the geographical area clustering based on the location context by semi-supervised learning algorithms. We perform spectral clustering on the locations based on their similarity of human activity distribution $A_{n,m}$. Each location of the grid has a distribution of activity categories with relative frequency of their appearance. The spectral clustering with $k$-nearest neighbor ($k=10$ based on cosine similarity metrics) approach allows us to classify geographical areas $L$ based on such multi-dimensional features, $A_{n,m}$. We then observed significantly different six types of areas, that are geo-located in Figure \ref{gra:clustering high level k =6}. The average values of the activity categories for those area types are presented in Figure \ref{gra:Activity vectors, clustering by k = 6}.

\begin{figure}[htb!]
       \centering
       \includegraphics[scale=0.35]{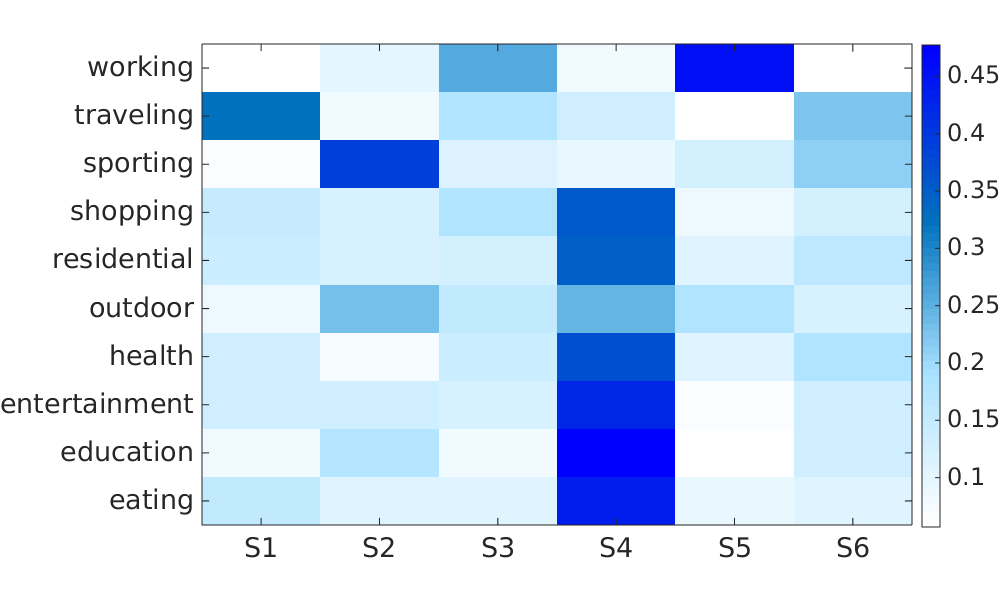}
       \caption{The average values of the activity categories in categorical area types observed}
       \label{gra:Activity vectors, clustering by k = 6}
\end{figure}


The figure shows that categorical area type $S4$ contains high percentage values for residential, and eating activities. The center of the city including a residential zone were clustered into one area type. The area type $S3$ contains high percentage value on working activity. This classification can be refined if we increase the number of area types observations. For each area type, we are now able to extract and observe timelines $T_{n,p,d}$ from mobile phone data records in order to determine the correlation between the timelines and the area profiles for those area types.

\begin{figure}[htb!]
       \centering
       \includegraphics[scale=0.35]{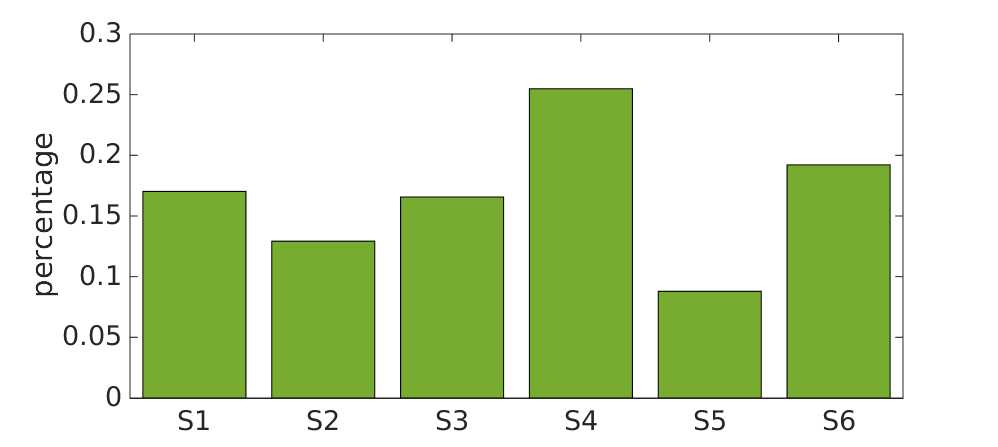}
       \caption{The density distribution of area types observed in Milan}
       \label{gra:cluster density}
\end{figure}

\begin{figure}[htb!]
  \centering
  \includegraphics[scale=0.45]{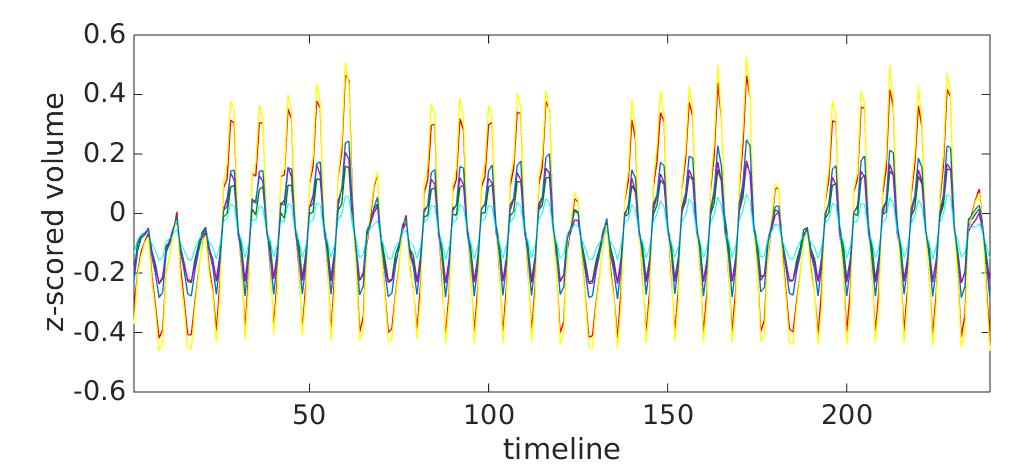}
  \caption{The average timeline of mobile phone data by area types ($k$=6), where $S1$ is red, $S2$ is lime, $S3$ is blue, $S4$ is yellow, $S5$ is cyan/aqua, and $S6$ is magenta/fuchsia}
  \label{gra:timelines hourly}
\end{figure}


%

The density of the clusters are almost uniform distributed except cluster S4 and S5, see Figure \ref{gra:cluster density}. This unbalanced datasets for clusters could contribute to an acceptable global accuracy, but also to a (hidden) poor prediction for instances in minority classes. In this context, alternative metrics, such as per class accuracy will be considered. We estimate the accuracy per class using the two techniques (canonical correlation coefficients vs learning techniques). Figure \ref{gra:timelines hourly} shows the actual volume of the mobile network traffic activities by the area types.

\begin{figure}[htb!]
       \centering
       \includegraphics[scale=0.4]{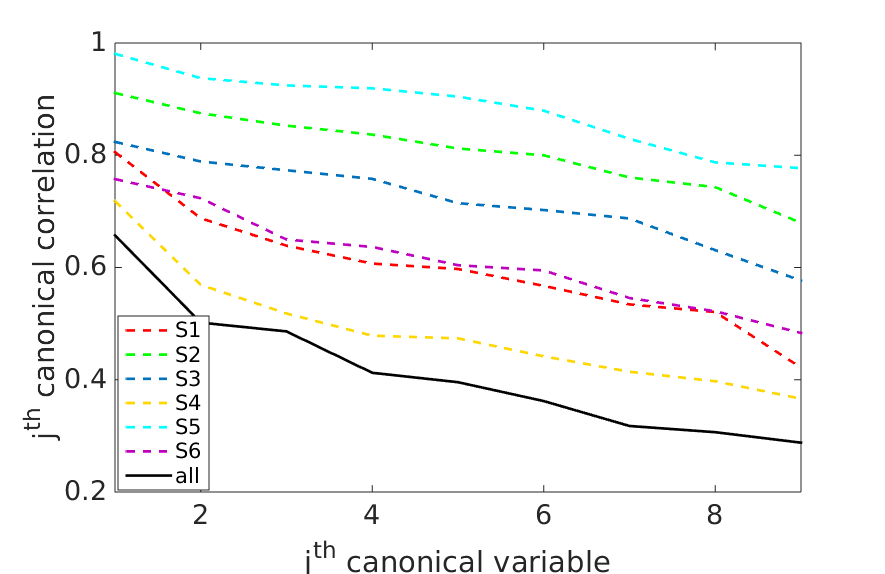}
       \caption{Canonical correlation between the two feature matrices for locations}
       \label{gra:area canonical coef}
\end{figure}

We illustrated the correlation between the area profiles $A_{n,m}$ and timelines $T_{n,p,d}$ based on the canonical correlation \cite{CanonicalAnalysis} (see Figure \ref{gra:area canonical coef}). The canonical correlation investigates the relationships between two sets of vectors by maximizing the correlation in linear combination. In order words, canonical correlation finds the optimal coordinate system for correlation analysis and the eigenvectors defines the coordinate system. While the overall maximum correlation coefficient ($j$=1) is 65\% between the two vectors, the correlation coefficient by area types is high between 72\% and 98\%. For example, the correlation in area type $S5$ is stronger than other area types, in which working type of activities are more distributed. The maximum correlation in $S2$ containing high percentage of sporting activity is 82.38\%.


We also compared the distance between the two vectors (mean) of area types to investigate the similarity of the relevant area profiles can have the similar human behaviors. We observed linear correlation with a coefficient of $r$ = 0.61 This result shows that as the distance between the area profiles is increased, the timeline difference increases, and human behaviors are strongly correlated to geographical area profiles.

In second, we profiles the communication timelines with the cluster labels observed in each location that will be used to estimate the correlation by supervised learning algorithms. The prediction accuracy of timeline types in a given location could be an evaluation of the dataset. To that end, we train several predictive models (i.e., Bayesian algorithms, Decision Trees, Probabilistic Discriminative models and Kernel machines.) to measure the prediction accuracy by k-fold cross validation method ($k$=10), which is used to estimate how accurately a predictive model will perform. We need to prepare training and test data. The training data are the timelines labeled by area types through the location. This allows us to determine if timelines are clustered as geographical area profiles. The experimental results on our data are shown in Table \ref{table:algorithms}. This classification of the predictive models is aimed at choosing a statistical predictive algorithm to fit in our analysis. 

\begin{table}[htb!]
\begin{center}
    \caption{Results for the predictive models with the use of area types observed by spectral clustering algorithm}
     \label{table:algorithms}
  \begin{tabular}{ | l | l | l |}
    \hline
    Algorithm & Cross Validation & Overall ACC \\ \hline
    Random classifier & 0.83 & 16.7\% \\ \hline
    Linear Discriminant & 0.5404 & 45.01\% \\ \hline
    Quadratic Discriminant & 0.4649 & 52.90\% \\ \hline
    Naive Bayes(kernel density) & 0.6748 & 20.38\% \\ \hline   
    K-NN (k=5, euclidean dist)  & 0.3822 & 61.73\% \\ \hline
    K-NN (k=10, euclidean dist)  & 0.4068 & 59.26\%\\ \hline
    Decision Tree & 0.4806 & 52.58\% \\ \hline
    Random Forest & 0.3513 & 64.89\%  \\ \hline
    Multi-class SVM & 0.4997 & 49.47\% \\ \hline
  \end{tabular}
\end{center}
\end{table}

\begin{figure}[htb!]
  \centering
  \includegraphics[scale=0.35]{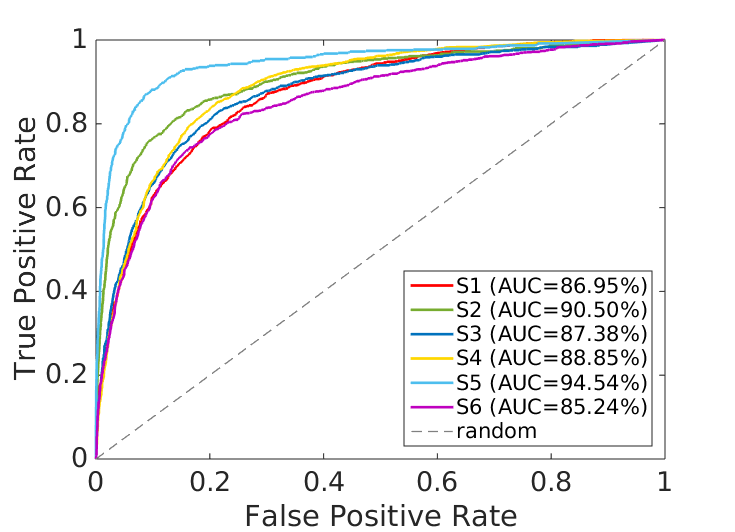}
  \caption{Receiver operating characteristic to multi-class by Random Forest classifier}
  \label{gra:roc}
\end{figure}

Among the considered techniques, the Random Forest and the Nearest Neighbor algorithms are resulted in the lowest error with high accuracy, in other words, if we take the area profile of the nearest-neighbor (the most common area profile of k-nearest-neighbors), that would give the right timeline type. The confusion matrix of the Random Forest classifier, and the precision, recall are estimated in the following Table \ref{table:confusion matrix1}. The receiver operating characteristic curve for visualizing the performance of the classifiers is described in Figure \ref{gra:roc}. This result shows that the area type S5 is the well classified and compact by showing a strong correlation between the area activity categories and area timeline. The area types S1, S2, S3 and S4, S6 can be still refined in terms of the area activity categories.

\begin{table*}[htb]
\begin{center}
    \caption{Confusion matrix and precision, recall and f-measure in each area type defined for predicting timeline based on location context about categorical human activity by Random Forest classifier}
     \label{table:confusion matrix1}
	\begin{adjustbox}{max width=\textwidth}
  \begin{tabular}{ | l | l | l | l | l | l| l|}
    \hline
    Area type defined & S1 & S2 & S3 & S4 & S5 & S6 \\ \hline \hline
S1 & 8.91\% & 0.20\% & 1.80\% & 4.47\% & 0.07\% & 1.57\%\\ \hline
S2 & 0.10\% & 8.58\% & 0.70\% & 1.77\% & 0.47\% & 1.30\%\\ \hline
S3 & 1.77\% & 0.43\% & 10.15\% & 1.70\% & 1.27\% & 1.23\%\\ \hline
S4 & 2.34\% & 0.53\% & 1.13\% & 19.63\% & 0.33\% & 1.54\%\\ \hline
S5 & 0.03\% & 0.23\% & 1.37\% & 0.53\% & 6.54\% & 0.07\%\\ \hline
S6 & 2.40\% & 1.27\% & 1.67\% & 2.74\% & 0.07\% & 11.08\%\\ \hline
  \end{tabular}
  \quad
    \begin{tabular}{ | l | l | l |}
    \hline
    Prec. & Recall. & F-measure. \\ \hline \hline
52.35\% & 57.30\% & 54.71\%  \\\hline
66.41\% & 76.26\% & 70.99\%  \\\hline
61.29\% & 60.32\% & 60.80\%  \\\hline
76.96\% & 63.64\% & 69.67\%  \\\hline
74.52\% & 74.81\% & 74.67\%  \\\hline
57.64\% & 66.00\% & 61.54\%  \\\hline
  \end{tabular}
  \end{adjustbox}
\end{center}
\end{table*}

%
%

\subsection{Land-use type vs human behavior}

\begin{figure*}[htb!]
       \centering
        \subfigure[The distribution of land-use classification in Milan]
       {\includegraphics[scale=0.18]{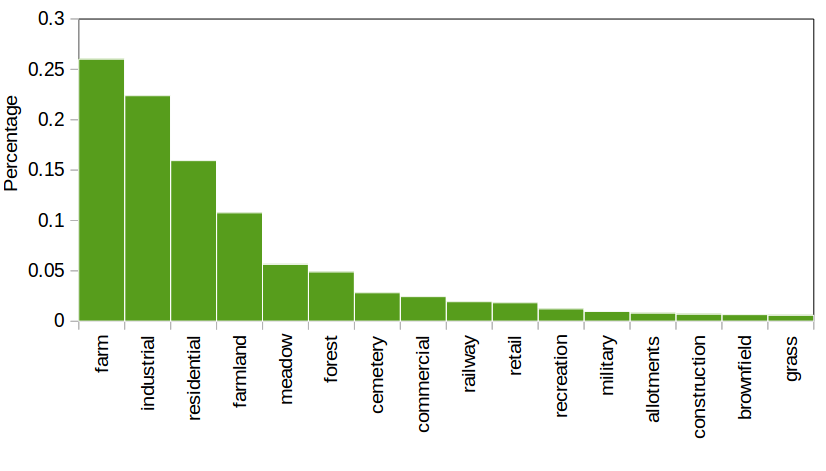}
       \label{gra:land-use management}}
       \quad
       \subfigure[The distribution of categorical activity clusters within commercial land-use area]
       {\includegraphics[scale=0.18]{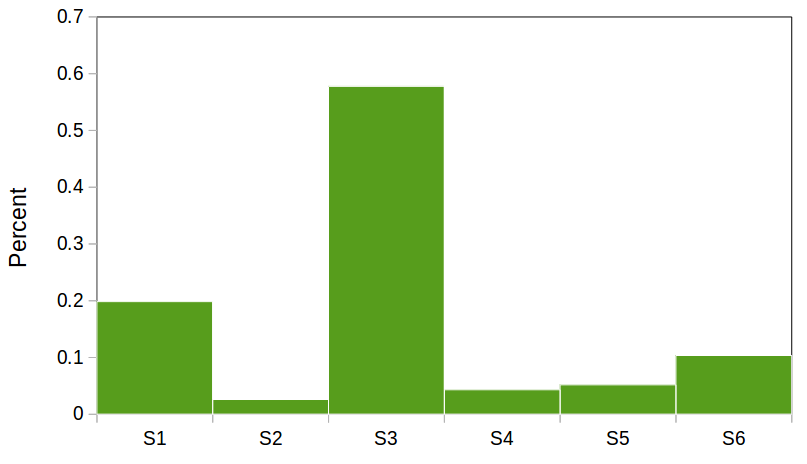}
         \label{gra:timelines commercial match}}
       \caption{}
\end{figure*}

After we observed strong prediction accuracy of timelines based on categorical area types, we analyze the relation between the timelines and land-use types which are formally defined by land-use management organizations. While many works try to predict activity based on land-use, we perform a comparative study of the two approaches. We identify that even the area of the same land-use might have different area types in terms of area profiles and those are still different in terms of human activity timelines quantified through mobile phone records, which validates significance of activity-based classification vs official land-use. We predict the timeline type of a given area based on the land-use type using the Random Forest and the Nearest Neighbor classifiers. We used the land-use types from the OSM\footnote{http:\///wiki.openstreetmap.org\//wiki\//Key:landuse} for this prediction task (see the distribution of land-use types of Milan in Figure \ref{gra:land-use management}). The prediction accuracy of the Random Forest classifier is 53.47\%. This shows that predicting power of categorical types is higher compared to land-use types.

\begin{figure}[htb!]
  \centering

  \includegraphics[scale=1,width=12cm, height=5cm]{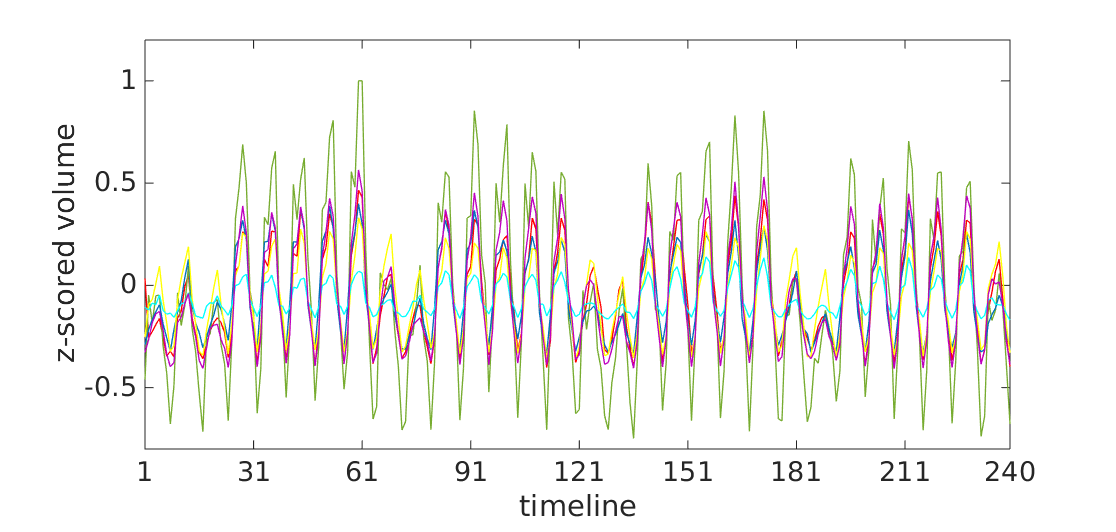}
    \hspace*{20cm}
  \caption{The timelines belong to different clusters within the commercial land-use: $S1$ is red, $S2$ is lime, $S3$ is blue, $S4$ is yellow, $S5$ is cyan/aqua, and $S6$ is magenta/fuchsia}
  \label{gra:timelines commercial timeline}
\end{figure}

We also match the area types we observed with the land-use types given officially. The result shows that even within the same land-use type, the timelines corresponding to different clusters are still different. For example, 58\% of the commercial land-uses matched with the area type S3 which followed by S1, S6 and S2, S3, S4, S5, shown in Figure \ref{gra:timelines commercial match}. The corresponding timelines to the different clusters within the commercial land-use type are illustrated in Figure \ref{gra:timelines commercial timeline}.

%
%
%
%
%

The timelines in the same area type observed, also in the same land-use officially defined, can be still refined, but the timeline pattern refinement will require more emphasis on the appropriate features, for example, timelines for weekday or weekend. The area profiles are semantically different concepts in terms of human activities performed in geographical areas. Further, it will allow us to identify a standard or exceptional type of mobile network activities in relevant areas, as well as to enable the identification of unknown correlations, or hidden patterns about anomalous behaviors.

\section{Conclusion and Future works}\label{sec:conclusion}

In this paper, we proposed an approach that characterizes and classifies geographical areas based on their anticipated (through POI distribution) human activity categorical types, such as working or shopping oriented areas. We concentrated on the analysis of the relationship between such spatial context of the area and observed human activity. Our approach compares the similarity between area activity categorical profiles and human activity timeline categories estimated through cell phone data records. We found an overall correlation of 61\% and canonical correlation of 65\% between contextual and timeline-based classifications. We observed six types of areas according to the area activity categories where we compared their human activity timelines with their area activity categories and the correlation (canonical) coefficient is between 72\% and 98\%. For example, the area type $S5$ related to working activity has a strong correlation of 98\% which followed by the area types, $S2$ related to sporting activity and $S3$ related to the human activities in the center of the city . The supervised learning approach validates possibility of using an area categorical profile in order to predict to some extent the network activity timeline (i.e., call, sms, and internet). For example, the Random Forest approach performs well with the accuracy of 64.89\%. So human behaviors' temporal variation is characterized similarly in relevant areas, which are identified based on the categories of human activity performed in those locations. Furthermore we found that the prediction accuracy based on the official land-use types is only 53.47\%. So the official land-use types by themselves are not enough to explain the observed impact  of area context on human activity timelines, also because even within the same land-use type, different activity categorical types still demonstrate different activity timelines. Further, the semantic description of area profiles associated to mobile phone data enables the investigation of interesting behavioral patterns, unknown correlations, and hidden behaviors in relevant areas. We expect the approach to be further applicable to other ubiquitous data sources, like geo-localized tweets, foursquare data, bank card transactions or the geo-temporal logs of any other service.

\section{Acknowledgments}
The authors would like to thank the Semantic Innovation Knowledge Lab - Telecom Italia  for publicly sharing the mobile phone data records which were provided for Big Data Challenge organized in 2013, Italy. We also would like to thank MIT SENSEable City Lab Consortium partially for supporting the research. 

\bibliographystyle{abbrv}
\bibliography{ref}

\end{document}